\documentclass[10pt]{article}

\usepackage{amsmath}
\usepackage{amssymb}

\usepackage{graphicx}

\usepackage{cite}

\usepackage{color} 

\usepackage[labelfont=bf,labelsep=period,justification=raggedright]{caption}

\makeatletter
\renewcommand{\@biblabel}[1]{\quad#1.}
\makeatother

\date{}

\begin{document}

% Title must be 150 characters or less
\begin{flushleft}
{\Large
\textbf{Earthquake Model Confirms Traffic Jams Caused by Tiredness}
}
% Insert Author names, affiliations and corresponding author email.
\\
Ferenc J\'arai-Szab\'o$^{1,\ast}$, 
Zolt\'an N\'eda$^{1}$
\\
\bf{1} Department of Physics, Babes-Bolyai University, Cluj-Napoca, Romania
\\
$\ast$ E-mail: jferenc@phys.ubbcluj.ro
\end{flushleft}

% Please keep the abstract between 250 and 300 words
\section*{Abstract}
A simple one-dimensional spring-block model elaborated for the idealized single-lane highway traffic reveals the causes for the emergence of traffic jams. Based on the stop-time statistics of one car in the row, an order parameter is defined and studied. By extensive computer simulations, the parameter space of the model is explored, analyzed and interpreted. Existence of a free a and congested flow phases is confirmed and the transition between them is analyzed.

% Please keep the Author Summary between 150 and 200 words
% Use first person. PLoS ONE authors please skip this step. 
% Author Summary not valid for PLoS ONE submissions.   
%\section*{Author Summary}

\section*{Introduction}

The rapid growth of the vehicle number during the last century caused an increased complexity in our road traffic and transportation systems. Unfortunately, in such conditions, traffic congestion becomes an everyday problem for drivers. The variety of complex non-linear phenomena present in such agglomerated traffic systems has attracted the attention of a large number of researchers (for a review please consult the recent work of Nagatani \cite{Nagatani2002}, or  Kerner's book \cite{Kerner2004}). Accordingly, since the early '30 many empirical data on different highways have been collected \cite{Greenshields1935,Kerner1997,Banks2002} and traffic data collection systems have been substantially evolved \cite{Atluri2009,Sifuentes2011}. On the theoretical side, many traffic models have been developed. The models are usually classified into four categories: microscopic models, macroscopic models, cellular automata models and non-traditional models. Detailed description and analysis of such traffic models can be found in many review articles \cite{Chowdhury2000,Helbing2001,Maerivoet2005,Darbha2008}. One recent example of a non-traditional stochastic model could be the probabilistic traffic flow theory \cite{Mahnke2005} that includes the nonlinear effects of small perturbations. It was also shown \cite{Treiber2006} that finite reaction times are clearly essential factors of driver behavior that affects the performance and stability of traffic. Despite of many existing studies in the field, the phenomenon of spontaneous traffic jam formation is far from a complete understanding. 

The simplest, but already quite complex form of the traffic is the accident-free and single-lane motion of a chain of cars. The motion of the queue in this simple form of traffic is primarily governed by the leading car and the statistics of driving attitudes. From empirical observations we know that this kind of traffic  can be either "free" with a continuous flow structure or congested with a stop-and-go motion of cars. The congested traffic usually appears if the leading car is moving slowly and the differences between the driving attitudes of drivers are substantial. In such situations the car row will evolve non-continuously in avalanches of widely different sizes, and a sequence of jams of different magnitudes will appear and propagate backward through the system. This form of traffic is also known as shockwave traffic jam or 'phantom traffic jam'. A systematic analysis of empirical traffic states of a 30 kilometer long section of the German freeway A5 near Frankfurt \cite{Schonhof2007,Schonhof2009} reported a rich variety of congested traffic states, interpreted as a spatial coexistence of altogether six different traffic states.

``Why do traffic jams appear on highways?'' - is the natural question asked by the authors of the first experimental article that investigates the spontaneous shockwave traffic jam formation \cite{Sugiyama2008,Nakayama2009}.  By a planned experiment performed on a circuit they show that the emergence of a traffic jam may occur even in the absence of a bottleneck. If one looks on their experiments (i.e. their video recording posted on youtube \cite{SugiyamaYoutube}) it is immediately observable that at the initial stage, vehicles are running continuously each with the same velocity, but roughly 10 min later a first shockwave traffic jam emerges spontaneously and propagates in the system. Shortly after this experiment these emergent shockwaves have been identified as nonlinear traveling wave solutions called 'jamiton solutions' of the purely deterministic hyperbolic continuum traffic equations \cite{Flynn2009}.

The present work is inspired and motivated by the experiments of Sugiyama et al. Based on their work, the formation of a traffic jam is understood as a collective phenomena. For a physicist, the most straightforward way to approach this phenomenon is by using the simple spring-block chain with asymmetric interactions, where block will model the cars and the springs the distance keeping tendency of drivers. Using this simple approach proved to be successful in modeling other collective phenomena such as earthquakes, fracture, fragmentation or even magnetization processes.  Here, through large-scale computer simulations on  simple model system the minimum conditions that are absolutely necessary to produce the self-organized congested traffic jam conditions are identified. Contrary to most of the studies in the field of highway traffic we focus on a measure characteristic for one car (block) in the row: the distribution of the rest times. Based on these distributions an order parameter will be defined and the parameter space of the model is explored and analyzed. It is shown that the transition between the free and congested traffic states is realized through a disorder-induced second order phase transition. Based on evident analogies with the model elements, it is concluded that besides the disorder level in driver's driving attitude the tiredness of drivers plays also an important role in the formation of traffic jams. 

\section*{Methods}

The spring-block approach of the single-lane highway traffic was described in detail in our previous work \cite{Jarai2011}. However, for the sake of completeness 
 a brief review of the model will be included here. 

\subsection*{From earthquakes to traffic modeling}

The used model belongs to a model family with broad interdisciplinary applications called spring-block type models. This model family was introduced in 1967 by R. Burridge and L. Knopoff \cite{Burridge1967} to explain the empirical law of Guttenberg and Richter \cite{Gutenberg1956} regarding the size distribution of earthquakes.
It consists of simple elements: blocks that can slide with friction on a horizontal plane connected in a lattice-like topology by springs. Later, due to the spectacular development of computers and computational techniques, this simple model proved to be very useful in describing many phenomena in different areas of science. Most of the collective phenomena that occur on mesoscopic scale in solid materials such as cracking or fragmentation in drying granular materials \cite{Leung2000,Leung2001}, capillarity driven self-organization \cite{Jarai2005,Jarai2007,Jarai2011b} or magnetization processes and Barkhausen noise \cite{Kovacs2007} can be modeled by spring-block type models. From these studies we learn that the model is always efficient when one deals with avalanches, complex dynamics or structure formation by collective behavior.

\begin{figure}[!ht]
\begin{center}
\includegraphics[width=8.3cm]{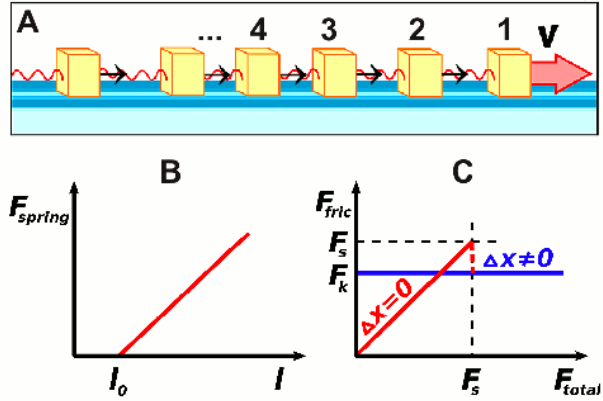}
\end{center}
\caption{
{\bf The spring-block model for highway traffic.} The used spring-block chain is shown on panel (a) while the force profile for spring-tension and friction are shown on panels (b) and (c), respectively.
}
\label{fig1}
\end{figure}

In the spring-block approach the idealized car queue that moves on a single highway lane is modeled  as a chain of blocks dragged by the first block that advances with a constant velocity $v$ (as sketched in the Figure \ref{fig1}A). Blocks are labeled after their ordinal number in the row so that the externally dragged block (the first in the queue) has label 1 and the last blocks in the queue is labeled with $N$. The position of the blocks are are denoted by $x^{(i)}$, where $i = \overline{1,N}$.

The blocks are connected by springs of equilibrium length $l_0$ and spring constant $k$ having the classical force profile shown in Figure \ref{fig1}B. The interactions resulting from the springs model the aim of drivers to keep a certain distance from the car ahead. Therefore, in order to be realistic, these springs cannot be classical mechanical springs with forces acting on both their ends, because in case of accident-free traffic the front car should not be influenced by the car from behind. Therefore, this distance keeping interaction acts only on the car in behind. In such system, the action-reaction principle is violated and, in this sense, it cannot be anymore considered as a realistic mechanical system.

It was recently shown, that in case of modeling the drivers' behavior, the reaction time, the velocity adaptation time and the numerical update time has to be considered \cite{Kesting2008}.  The last two of these characteristic times will be included in the dynamics of the spring-block model presented later. Another important ingredient which can be incorporated in the model is the disorder generated by the driving style of drivers. From our point of view, this should be the drivers' inertia which indicates how quickly the driver can react to a certain event or how quickly can follow the velocity change of the car ahead (reaction time). In order to keep our model as simple as possible, this is introduced via friction forces between the blocks and the plane on which they slide. These friction forces are generated randomly and independently for each new position of the blocks. As a first approximation we consider for this quantity a normal distribution with a fixed mean $\langle F_s \rangle$ and standard deviation $\sigma$.  As a result of this both a spatial and temporal disorder is introduced. Spatial disorder means differences in driving styles while temporal disorder means fluctuations of driving attitudes in time. The mean value of the friction force may be connected to the average reaction time of the drivers. Short reaction times are expected in case of fresh, rested drivers. As the driver get tired, these reaction times are gradually increased.

In analogy with real mechanical systems, static and kinetic friction forces are considered. They are denoted by $F_s$ and $F_k$, respectively. The friction force profile is presented in Figure \ref{fig1}C. The selection of the friction force value depends on the previous displacement $\Delta x$ of the block and the total force $F_{total}$ acting on it. Similarly with classical dynamical systems, the maximum possible value of the static force is considered to be greater than the kinetic one. For simplicity 
reasons in our model their ratio is kept constant $f = F_k / F_s$.  By this assumption we include in the model the tendency of drivers to react quicker to an event when the car is in motion. In such cases, there is no need to release the breaks and then, step the acceleration and slowly release the clutch as would be the case for a stopped car. However, in case of cars with automatic transmission this difference may be smaller which leads to $f$ ratio values closer to $1$.

The simulation follows the typical steps of a simplified molecular dynamics simulation and is summarized bellow. 

\begin{itemize}
\item \emph{Simulation step \#1: calculation of forces acting on each block.} All blocks are visited and the spring force $F_{spring}$ acting on each block $i$ 
is calculated by
\begin{equation}F_{spring}^{(i)} (t) = k  \left[ l^{(i)}(t-1) - l_0 \right]\,,\end{equation}
where $l^{(i)}$ denotes the length of the spring connected to the front of the i-th block. If the block is at rest (its previous displacement $\Delta x^{(i)}(t-1) = 0$) the spring force is compared to the static friction. If $F_{spring}^{(i)}(t) \leq F_s^{(i)}(t)$ the static friction equals the spring force and the total force acting on the block will be 
\begin{equation}F_t^{(i)} = 0\,.\end{equation}
Accordingly, the block will remain at rest in this step. On the contrary, if $F_{spring}^{(i)}(t) < F_s^{(i)}(t)$ the block will start to move and the kinetic friction force is added to the spring force resulting:
\begin{equation}F_t^{(i)} = F_{spring}^{(i)} - F_k^{(i)}\,.\end{equation}
 The total force is similarly calculated, when the block is not at rest ($\Delta x^{(i)}(t-1) \neq 0$).

\item \emph{Simulation step \#2: calculation of block displacements.} All blocks are visited and, based on forces $F_t^{(i)}$ calculated in simulation step \#1, their displacements $\Delta x^{(i)}$ are calculated and stored. The displacement of the first block is $\Delta x^{(1)} = d_0$ which represents the constant drag step of the first block. For the rest of the blocks $i = \overline{2,N}$ the displacement is calculated by using the equation of classical mechanics with a properly selected timestep $\Delta t$ 
\begin{equation}\Delta x^{(i)}(t) = \Delta x^{(i)}(t-1)+ A \, F^{(i)}(t)\Delta t^2\,,\label{eq_displacement}\end{equation} 
where $A = 1 / 2m$, $m$ being the mass of a block. 

\item \emph{Simulation step \#3: application of restrictions.} For the calculated displacements, the block step limit $d_{max}$ is applied which means that all displacement values grater than $d_{max}$ are set to be $d_{max}$.

\item \emph{Simulation step \#4: updating block positions.} All blocks are visited following their ordinal numbers and their new positions are calculated by
\begin{equation}x^{(i)}(t) = x^{(i)}(t-1) + \Delta x^{(i)}(t)\,.\end{equation}
Here the only restriction is that the minimum distance $d_{min}$ between blocks has to be respected.

\item \emph{Simulation step \#5: data collection.} Stop times for the selected block are detected and stored in the output file. Stop time of a car is defined 
as the total time-length of consecutive simulation steps for which the displacement is zero. This is recorded at the time-moment where the car begins to move again. Car positions can also be stored at this step but this will result in a large amount of data. This feature is usually turned on for visualization purposes in case of small systems only.
\end{itemize}

Simulation steps \#1--5  are repeated until enough data is collected for a proper statistical analysis.

In our previous study \cite{Jarai2011} it was investigated how the position of a single block in the queue is influencing its stop-time distribution. From this study we learned that after a certain number of blocks (transient distance), their cumulative stop-time distribution converges to the same function. Therefore, in order to ensure that
the right asymptotic behavior is studied, in the present study we simulate block chains longer than this transient distance, and the statistics for the 
last block in the row is investigated. 

\subsection*{Model parameters and the reality}

The motion of the queue is simulated in discrete time-steps of length $\Delta t = 1$. This fixes the unit for simulation time and gives us the possibility to handle easily the stochasticity in the equations of motion. 

The length unit in the simulation is defined by the length of a single block $L=1$. This distance is not relevant for the studied statistics, but it has 
an importance in determining realistically some other relevant distances.  In order to simulate the simplest dynamics without accidents, a minimum distance between blocks $d_{min} = 0.3$ is imposed. This distance is considered to be also the equilibrium length $l_0$ of the asymmetric springs. Taking into account that the length of a real car is about $3-4$ m, this minimum distance corresponds to a $1$ m minimum following distance.
In the present simulations also a step limit of $d_{max} = 1$ is imposed for each car. This defines a speed-limit for the blocks inside the row.

The motion of the queue is governed by the drag step (the movement of the first block) which is kept constant in time. In each unit of time the first block moves ahead in steps of length $d_0$ which fixes the velocity of the first car. 

\begin{figure}[!t]
\begin{center}
\includegraphics[width=8.3cm]{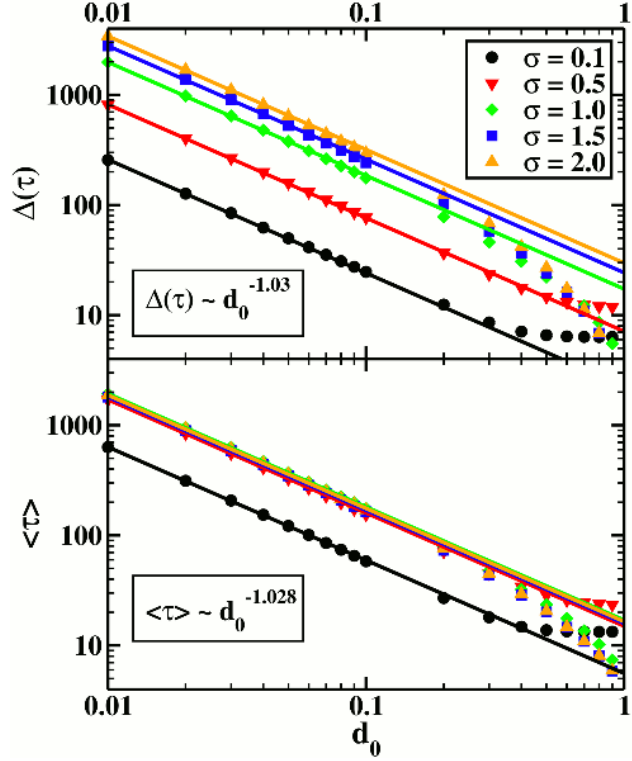}
\end{center}
\caption{
{\bf Scaling of characteristic stop-times with the drag-step.}  Stop-time averages (bottom) and standard deviations (top) as a function of the drag steps.  Results obtained from a statistics of 100000 stop-times. Simulation results for different $\sigma$ disorder level are presented. The quantities in this Figure are presented in simulation units. The simulation data is sourced from \cite{Jarai2011}.
}
\label{newfig2}
\end{figure}

In our previous study \cite{Jarai2011} it was shown, that except for high drag step values the average stop-time and the standard deviation of stop-times scales with the drag step following a $1/d_0$ functional dependence for any other model parameter values. In Figure \ref{newfig2} this scaling is repeated for different disorder levels 
$\sigma$. In interpreting this scaling feature one has to keep in mind that in these simulations the time-step is kept constant. Thus, in order to reach the continuum limit, the drag step has to be reduced down to a value where this scaling is valid. Moreover, for small drag step values $d_0 << 1$ the scaling shows that the behavior of the system is statistically the same for any small drag velocities. In other words, in our later investigations this parameter value should be fixed to a small value (i.e. $d_0 = 0.05$) and then the results and conclusion will be valid for any drag velocities that are much smaller than the speed limit of the blocks. More evidences in this sense will be presented later, in the Results section of the article.

The motion of each block is governed by the total force acting on it. The force unit in the model has been selected by considering that all springs have the spring constant $k = 1$ and unit elastic force acts on the block from behind when the distance between the two blocks is $d_{min}+1$ expressed in simulation length units. With this force unit the mean value of the normal distribution for friction forces $\langle F_s \rangle$ and the standard deviation $\sigma$ may be freely selected.

The constant $A$ present in the equation (\ref{eq_displacement}) is chosen to be $A = 1$ which fixes the mass of blocks to $m = 0.5$ expressed in simulation units.

Therefore, in the following study the effect of the remaining three freely adjustable model parameters on the system dynamics will be investigated. They are the ratio $f$ of the static and kinetic friction forces, the mean value of static friction force $\langle F_s \rangle$ and the disorder level $\sigma$. 

% Results and Discussion can be combined.
\section*{Results}

In real traffic conditions the data is usually collected by detectors located at fixed places. Accordingly, the number of passing cars in a given time interval (vehicle flow) is the most studied quantity. There are also other related and measured quantities like time headway, time clearance and time occupancy \cite{Helbing2001}. These quantities are, however relevant for the whole system and they are not focusing on a single car in the row. 

In our previous study \cite{Jarai2011} we turned the question around and we focused on what a driver stuck in the middle of a free or a congested traffic is experiencing. This can be quantified by the stop-time distribution $g(\tau)$ of the selected car in the row. This stop-time distribution describes the distribution of time intervals during which a given car is not moving. In other words, this distribution function $g(\tau)$ determines the probability density that the rest time of the car is $\tau$. It also gives a hint on how predictable the traffic flow is for the driver. For instance, if the stop-time distribution has a narrow peak, drivers can predict how much time they are blocked in a given position. On the contrary, for a power-law type long tail distribution the waiting times are widely different and thus unpredictable.

Concerning the dynamics of the studied spring-block system interesting and non trivial results have been reported in \cite{Jarai2011} by studying 
the  average stop-time values, and the standard deviation of the stop-time values. Also, interesting results were obtained while studying the influence of 
the disorder level on the friction force values. It was observed  that for a fixed mean static friction force value $\langle F_s \rangle$ at a certain disorder level ($\sigma \simeq 0.7$) there is a maximum in the average stop-time value \cite{Jarai2011}. This result suggested for example that in our simple traffic system there is a ``worst'' disorder level in driving attitudes (modeled by friction forces) for which the average stop-time of a block in the row is maximum. 
 
\subsection*{Disorder induced phase transition}

The disorder level in the system will be quantified in the following by dividing the stop-times standard deviation $\Delta (\tau)$ 
with the corresponding stop-time average $\langle\tau\rangle$.  In 
such manner,  one single dimensionless parameter is obtained 
\begin{equation} r = \frac{\Delta (\tau)}{\langle\tau\rangle}\,,\end{equation} 
which is appropriate as an order parameter. This order parameter will characterize the relative disorder in waiting times experienced by a driver.
The simulations performed in the present work will focus on computing this order parameter 
as a function of the relevant $f$, $\langle F_s \rangle$ and $\sigma$ parameters of the model. 

\begin{figure}[!tb]
\begin{center}
\includegraphics[width=6.5cm]{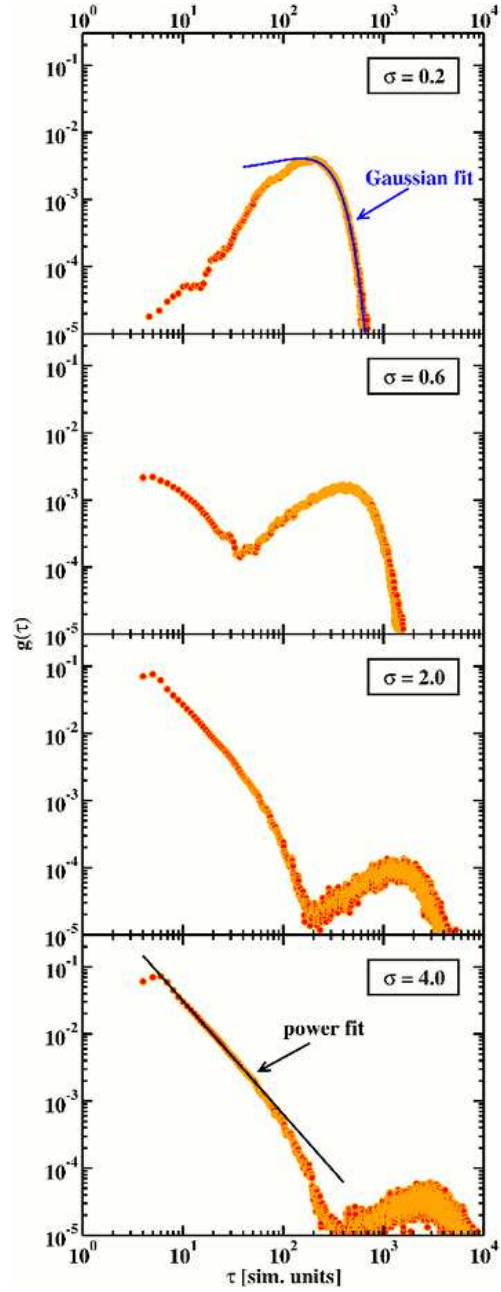}
\end{center}
\caption{
{\bf Characteristic stop-time distributions.} Normalized stop-time distributions for the last block in the chain for different disorder levels in the friction force $\sigma$. Results
obtained from a statistics of 100000 stop-times.
}
\label{fig2}
\end{figure}

As an example, in the following a case study will be performed for fixed $\langle F_s \rangle = 4$ and $f=0.8$ parameter values. 
Stop-time distribution functions $g(\tau)$ are constructed from 100\,000 simulated stop-times and they are plotted in the  Figure \ref{fig2} for different disorder levels 
$\sigma$.  As it is immediately observable, near the disorder level $\sigma = 0.6$ (second panel on the Figure \ref{fig2}) there are
two peaks with the same magnitude in the stop-time distribution function. At lower disorder levels (first panel of Figure \ref{fig2}) the first peak disappears and the stop-time distribution of long stop-times will be close to a normal distribution. This may be explained by assuming that in this region the motion of blocks is independent of each other and it is mainly influenced by friction forces generated randomly from a normal distribution. At higher disorder levels however (third and fourth panel of Figure \ref{fig2}) a power-law type distribution forms, suggesting correlated stops induced by collective motion of blocks. 

\begin{figure}[!ht]
\begin{center}
\includegraphics[width=16cm]{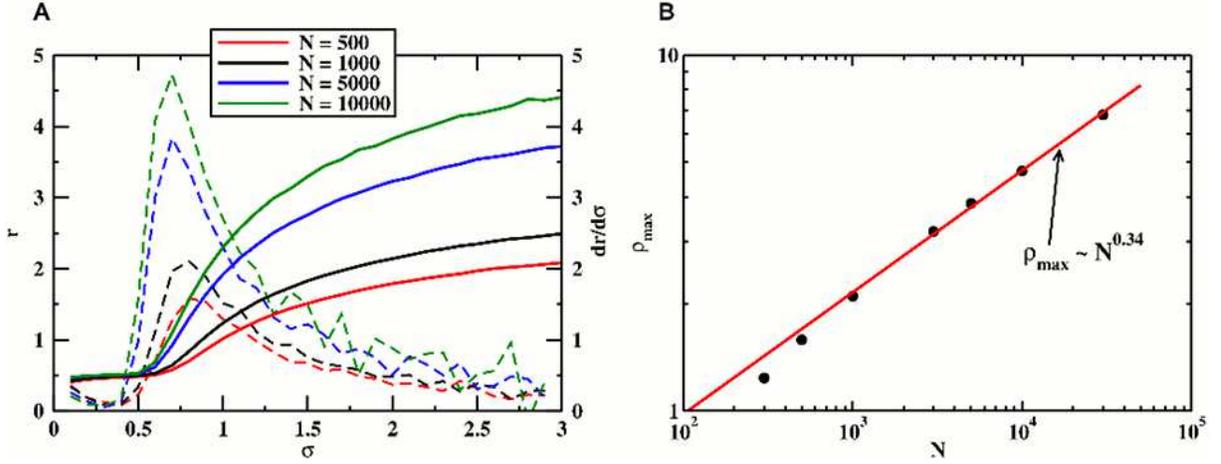}
\end{center}
\caption{
{\bf Finite size effects.} The order parameter $r$ (continuous lines) and its first derivative $\frac{dr}{d\sigma}$ (dashed lines) as a function of the disorder level in the system $\sigma$ are shown on the panel A. Results for different system sizes, $N$, are plotted. On the panel B the maximum of the first order derivative $\rho_{max}$ is plotted as a function of the system size. The parameter values for the simulations are the same as in Figure \ref{fig2} and all the quantities are expressed in simulation units.
}
\label{fig3}
\end{figure}

Let us visualize the above results using the order-parameter $r$ calculated from the same statistical data.  Results  as a function of the disorder level $\sigma$ are plotted with continuous lines in the Figure \ref{fig3}A. For investigating finite-size effects systems with different sizes are also considered. The graphs suggests that the order parameter has small values around $r \approx 0.5$ for low disorder levels ($\sigma \in [0.1, 0.7]$) and for higher disorder levels the order parameter is quickly reaching values grater than $1$. Therefore, as a function of the disorder level in the system two different phases may be identified: the free flow phase and the self-organized congested flow (or jam) phase. The transition between these modes is realized through a disorder induced  athermal phase-transition. As the system size is increased steeper transition regimes are observable, in agreement with what one would expect for a second-order phase-transition. Plotting the derivative of the order parameter $\frac{dr}{d\sigma}$ for different system sizes (dashed lines on the Figure \ref{fig3}A) confirms the above conjecture. As plotted on the Figure \ref{fig3}B the maximum of the first order derivative $\rho_{max}$ around $\sigma \approx 0.8$ is increasing  as a power law with the system size, indicating that the first derivative of the order parameter diverges in case of infinite systems. This is a clear sign of a second order phase transition.

\begin{figure}[!ht]
\begin{center}
\includegraphics[width=8.3cm]{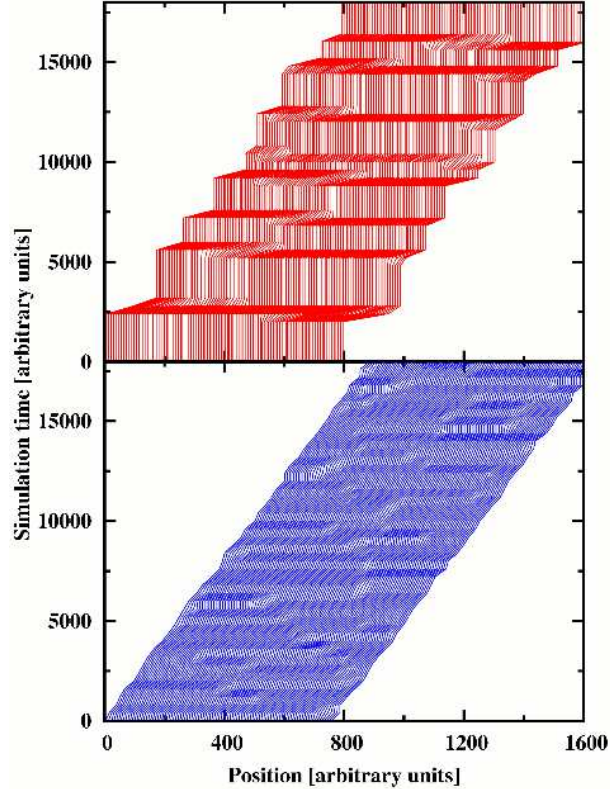}
\end{center}
\caption{
{\bf Time diagrams for the traffic of the cars.} Position of the last 50 cars (in a queue of 1000 cars) plotted in time for two different $\sigma$ values. The diagrams of block motions in the bottom panel indicates a continuous flow ($\sigma=0.2$). The diagrams in the top panel indicates a congested flow phase ($\sigma=3$).
}
\label{fig4}
\end{figure}

The motion of blocks in each phase of these two phases may be visualized by means of position--time diagrams. These are plotted in the Figure \ref{fig4}. On these
graphs the position of each block is plotted along the horizontal axis at each simulation time-step considering the last $50$ blocks of a row of $1000$ units. In the bottom panel characteristic simulation results for systems with low disorder level (here with $\sigma = 0.2$) are presented. One will observe here that except of small perturbations, the spatiotemporal pattern of an almost continuous flow is outlined with short stoptimes that are not visible at the scale of the graph. On the contrary, in case of high disorder levels (here $\sigma = 3.0$ is chosen), the spatio-temporal pattern (top panel of Figure \ref{fig4}) is stepwise with clearly visible stop-times on different time-scales and short headway periods between them. During this headway a self-organized shockwave-like motion of blocks may be detected. 

Therefore, our simulation results suggest that the dynamics of the studied spring-block system viewed from the reference-frame of one block exhibits non-trivial, critical behavior. A disorder induced, athermal second-order phase transition is observed, separating the independent block motions from a highly correlated avalanche-like collective motion.

\subsection*{Exploring the parameter space map}

\begin{figure}[!ht]
\begin{center}
\includegraphics[width=6cm]{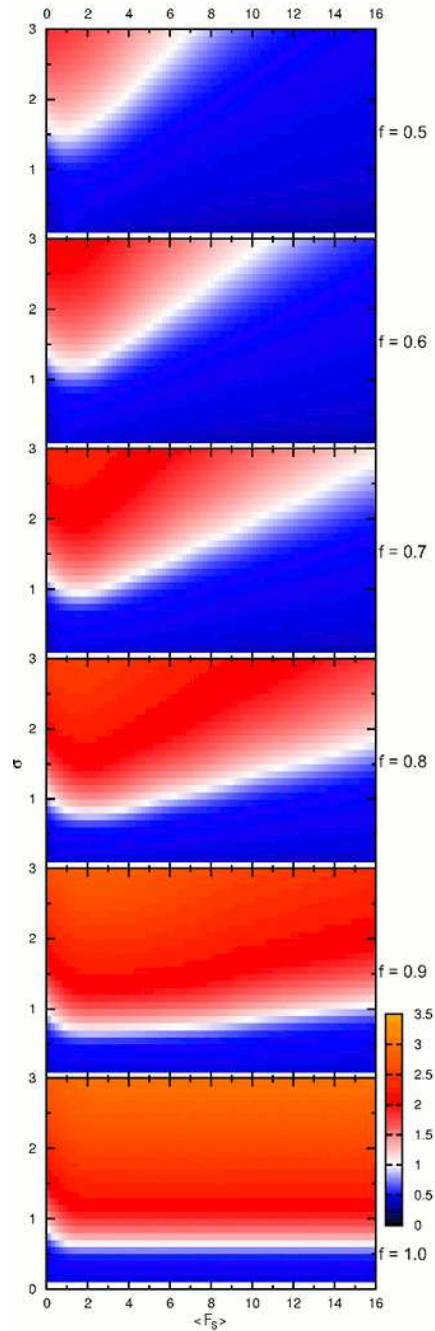}
\end{center}
\caption{
{\bf Mapping of the parameter space.} The color maps code the value of the order parameter, $r$, as a function of the $\langle F_s \rangle$ and $\sigma$ parameters expressed in simulation units. 
Different panels correspond to different $f$ ratios.  
}
\label{fig5}
\end{figure}

Finally, using the order parameter defined and analyzed in the previous section the relevant parameter-space of the spring-block traffic model is mapped. In six different color maps, on Figure \ref{fig5} the order parameter is represented as a function of the two main parameters of the model: the mean value of static friction force varied between $\langle F_s \rangle \in[ 0,  16]$ and the disorder level (varied between $\sigma \in [0.1,  3.0]$. Each created map is for different kinetic/static friction force ratio taken in the interval $f \in [0.5, 1.0]$. The value of this $f$ parameter is 
indicated near each map on these Figures.  Each map contains 2430 data points and each point is calculated from a statistics of 100\,000 simulated stop-times. The color coding has been calibrated in such way that blue color tones represent order parameter values in the range $r \in [0,0.7]$, red to orange color tones represent order parameter values in the range 
$r \in [1.2, 3.5]$. The transition regime $r \in [0.7, 1.2]$ is marked with different shades of gray and white. 

\begin{figure}[!ht]
\begin{center}
\includegraphics[width=16cm]{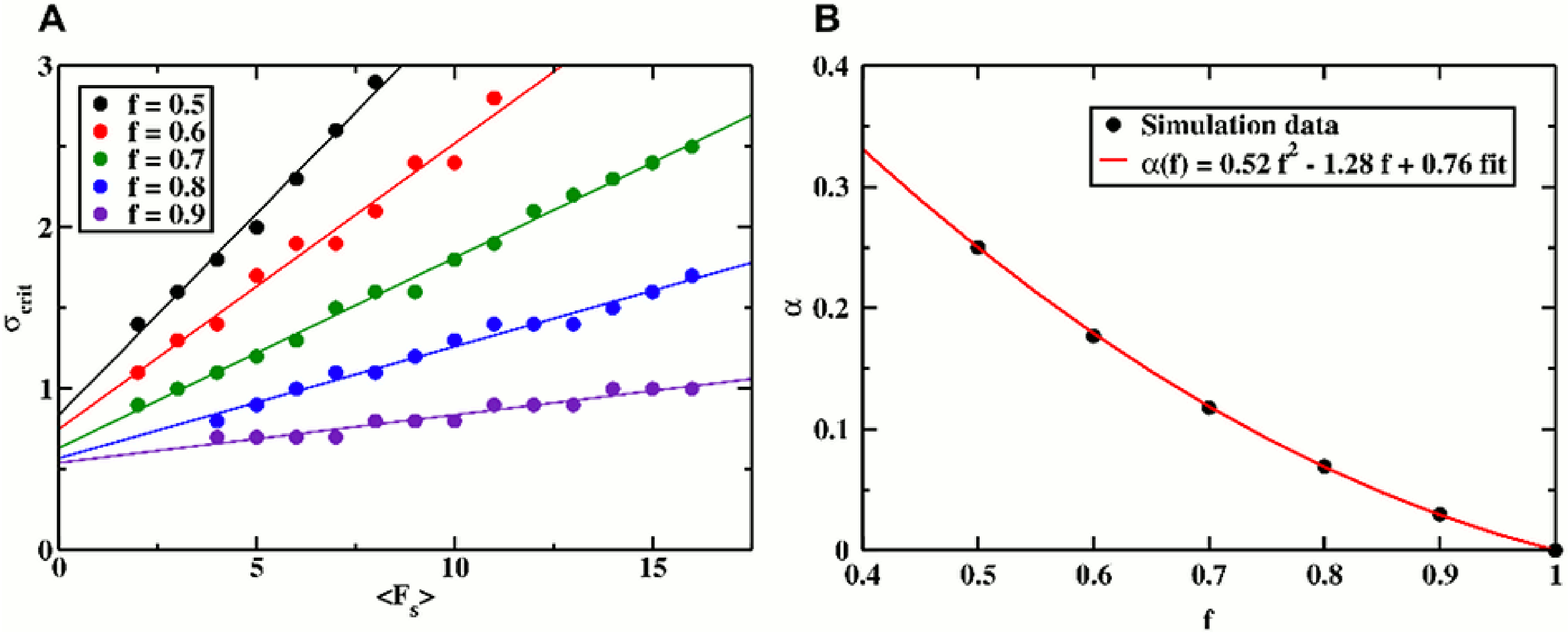}
\end{center}
\caption{
{\bf  Topology of phase-boundaries} Figure (a) presents the critical disorder level as a function of the $\langle F_s \rangle$ static friction value.  
Figure (b) plots the slope $\alpha$ of the $\sigma_{crit}-\langle F_s \rangle$ lines as a function of the kinetic to static friction force ratio, $f$. The quantities on both panels are expressed in simulation units.
}
\label{fig6}
\end{figure}

In agreement with the previous case study performed for a single $\langle F_s \rangle$ value, on each parameter map the free flow and jam phases are clearly separated.  For not too small $\langle F_s \rangle$ values the location of the phase separation point $\sigma_{crit}$  (determined from the inflection points of the $r-\sigma$ curves) is well approximated by a linear dependence (see Figure \ref{fig6}A):
\begin{equation}
	\sigma_{crit} (\langle F_s \rangle) = \alpha (f) \langle F_s \rangle + \beta\,.
\end{equation}
The slope $\alpha$ is decreasing as the $f$ value increases up to 1.0 according to quadratic trend (Figure \ref{fig6}B). The relations from above allow us to determine the disorder level $\sigma_{crit}$ for any combination of the $f$ and $\langle F_s \rangle$ parameter values. 

\section*{Discussion}

\begin{figure}[!ht]
\begin{center}
\includegraphics[width=8.3cm]{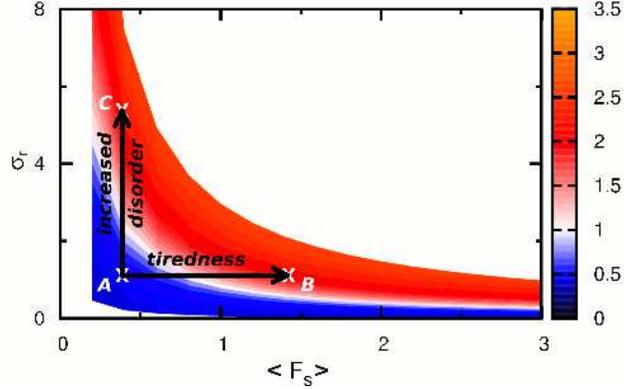}
\end{center}
\caption{
{\bf Effect of tiredness and differences in driving attitudes.}  The color-code indicates the order parameter value as a function of the $\sigma_r=\frac{\sigma}{\langle F_s \rangle}$ relative standard deviation and $\langle F_s \rangle$  values expressed in simulation units. Tiredness of drivers or increased disorder in driving attitudes drives the system from a continuous flow phase to a congested traffic phase. Phantom traffic jams are thus 
a result of either increased disorder in driving attitudes or increased tiredness of drivers. 
}
\label{fig7}
\end{figure}

Let us review now the experiment of Sugiyama et al. \cite{Sugiyama2008} in the viewpoint of the new modeling results. On their published video \cite{SugiyamaYoutube} vehicles are following each other on a circular track with the same imposed average velocity. After 10 minutes of circling, jams begin to emerge and propagate in the system. It is obvious, that after the long,  boring and monotonic course, the drivers' attention reduces and their reaction time increases \cite{Corfitsen1994}. In our spring-block approach this is equivalent to an increased average static friction value, $\langle F_s \rangle$. Assuming that the static friction value of all drivers and in all positions decreases proportionally and the micro-ensemble (characterized by the blocks relative friction values) is not changed, one gets that as $\langle F_s \rangle$ is increasing in the system the relative standard deviation $\sigma_r=\frac{\sigma}{\langle F_s \rangle}$ should remain constant. In order to have a more appropriate view on the behavior of the spring-block chain under such assumptions the order parameter map is reconstructed.  On Figure \ref{fig7} we present results in such sense for a single $f=0.8$ value. In contrast with figures \ref{fig5}, instead of the absolute standard deviation $\sigma$, the relative standard deviation $\sigma_r$ is considered on the vertical axes. In this representation, the increase in the drivers reaction time is equivalent with a horizontal displacement of the system's characteristic point to the left (i.e. from point A to B on the Figure). As indicated on the Figure, by such displacements it is possible to have a transition from the free flow phase (blue region) to the jammed phase (red region). Accordingly, our model system confirms that tiredness may induce the emergence of phantom traffic jams. The other way to have a transition from the free flow phase to the jammed flow phase (i.e. from point A to C on the Figure) is through an increased disorder in driving attitudes which may be induced by some unexpected traffic conditions.

\begin{figure}[!t]
\begin{center}
\includegraphics[width=8.3cm]{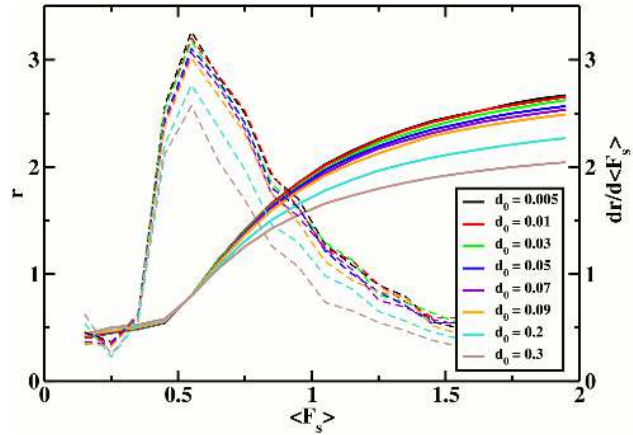}
\end{center}
\caption{
{\bf Effect of drag velocity on the obtained phase-transition.} The order parameter (continuous lines) and its first derivative (dashed lines) are plotted as a function of the static friction for different $d_0$ drag-step values. All quantities are expressed in simulation units. The results confirm that the obtained phases are the same independently of the chosen drag-step values. 
}
\label{fig8}
\end{figure}

It is important also to comment on the influence of the drag velocities or drag steps, $d_0$, on the observed statistics. Based on the simple scaling relation for the stop-time 
average and standard deviation with the drag step (Figure \ref{newfig2}), we already emphasized that  for small drag step values $d_0 << 1$ the system's behavior is re-scalable.  
This presumption may be confirmed performing a series of simulations where the relative standard deviation $\sigma_r$ and the kinetic/static friction force ratio $f$ is kept constant. Figure \ref{fig8} shows results in such sense. Here the order parameter and it's first derivative is plotted as a function of $\langle F_s \rangle$ for different drag step values, $d_0$. From 
the obtained curves it is immediately observable that in the small drag step limit the critical point separating the two phases is not influenced by the 
chosen drag step value. The phases that appear for different drag velocities are thus not influenced by the chosen drag velocity, assuming that this value is small. The obtained results are thus easily generalizable for all small drag velocity values.

In conclusion, a simple one-dimensional asymmetric spring-block model inspired from the early earthquake models has been elaborated for the idealized single-lane highway traffic.  The model has three free parameters, and a thorough mapping of the parameter space indicates two clearly distinguishable phases and a rich collective behavior. A disorder induced second-order phase transition has been observed, separating the free flow phase from the congested flow phase which is dominated by a highly correlated avalanche-like collective motion. Through these computer simulation studies it was confirmed that the emergence of phantom traffic jams characterized by continuous formation and propagation of jamitons [14, 17] may be induced by the tiredness of the drivers. Another effect that can lead to appearance of phantom traffic jams is the increased differences of driving attitudes. 

% You may title this section "Methods" or "Models". 
% "Models" is not a valid title for PLoS ONE authors. However, PLoS ONE
% authors may use "Analysis" 
%\section*{Materials and Methods}

% Do NOT remove this, even if you are not including acknowledgments
\section*{Acknowledgments}
This work was supported by Romanian National Research Council (CNCS-UEFISCSU), project number PN II-RU PD\_404/2010. The funders had no role in study design, data collection and analysis, decision to publish, or preparation of the manuscript. No additional external funding received for this study.

\section*{Author Contributions}
Wrote the paper: ZN FJS. Conceived the model: ZN FJS. Wrote the simulation program and performed the simulations: FJS. Interpreted the results: ZN FJS.

%\section*{References}
% The bibtex filename
\bibliography{plos-sb-traffic}

\end{document}